\documentclass[a4paper,11pt]{article}
\pdfoutput=1 

\usepackage{jcappub} 

\usepackage[T1]{fontenc} 

\usepackage{amsmath,amssymb,bm,mathrsfs}
\usepackage{url}
\usepackage{graphicx}
\usepackage{color}
\usepackage{mathtools}
\usepackage{lipsum}
\usepackage{cancel}
\usepackage[normalem]{ulem}

\title{\boldmath \fontsize{19pt}{19pt}\selectfont 
Pseudo-Nambu-Goldstone Boson Production \\
from Inflaton Coupling during Reheating}

\author[a]{\large Kunio Kaneta}
\author[b]{Sung Mook Lee}
\author[c]{Kin-ya Oda}
\author[d]{Tomo Takahashi}

\affiliation[a]{Faculty of Education, Niigata University, Niigata 950-2181, Japan}
\affiliation[b]{Department of Physics, Korea Advanced Institute of Science and Technology, Daejeon
34141, Korea}
\affiliation[c]{Department of Mathematics, Tokyo Woman’s Christian University,
Tokyo 167-8585, Japan}
\affiliation[d]{Department of Physics, Saga University,
Saga 840-8502, Japan}
\emailAdd{kaneta@ed.niigata-u.ac.jp}
\emailAdd{sungmook.lee@kaist.ac.kr}
\emailAdd{odakin@lab.twcu.ac.jp}
\emailAdd{tomot@cc.saga-u.ac.jp}

\abstract{
The existence of pseudo-Nambu-Goldstone boson (pNGB) fields is a common feature in many models beyond the Standard Model, characterized by their exclusive derivative couplings. This paper investigates a scenario where a pNGB is coupled to the inflaton field during the reheating phase of the early universe. We calculate the perturbative decay rate of a coherently oscillating inflaton into pNGBs on a general basis, considering both constant and field-dependent couplings with monomial potentials at the minimum. As a concrete application, we explore the production of axions when the radial mode of the Peccei-Quinn (PQ) scalar serves as the inflaton, particularly in the presence of a large gravitational non-minimal coupling. Our findings suggest that the presence of pNGBs during reheating can lead to significant non-thermal relics, offering new constraints on inflationary reheating models and providing potential observational signatures in the form of dark radiation.
}

\begin{document}
\maketitle
\flushbottom

\section{Introduction}

In the framework of inflationary cosmology, the beginning of the thermal universe is not simply marked by a `big bang' but by a decay and subsequent thermalization of the inflaton field that drives the exponential expansion during inflation \cite{Linde:1981mu,Albrecht:1982mp,Kofman:1994rk,Kofman:1997yn}.

After inflation, the inflaton field begins to oscillate coherently around the minimum of its potential, decaying into other fields in a process known as `reheating' (even though this is the \textit{first} instance of heating for our universe). Due to the coherence and time dependence of this process, non-perturbative, resonant particle production can occur in the early stages of reheating, a phase referred to as `preheating' \cite{Kofman:1997yn}. (For a comprehensive review, see \cite{Lozanov:2019jxc}.) Consequently, the reheating stage is critical for setting the initial conditions of the thermal universe. The coherence of this process also can lead to different equations of state, depending on the shape of the potential near its minimum.

In the standard slow-roll, single-field inflationary model, predictions are largely insensitive to the details of the reheating phase, thanks to Weinberg's theorem on the conservation of the adiabatic mode \cite{Weinberg:2003sw}. Eventually, the decay products are assumed to thermalize, erasing all information except for the reheating temperature $T_{\rm reh}$.\footnote{However, there can be indirect effects from the reheating stage, such as altering the number of $e$-folds during inflation \cite{Cook:2015vqa,Cheong:2021kyc}.}

This paper explores the possibility of remnants from the reheating stage, specifically non-thermal relics as the form of `dark radiation' resulting from inflaton decay. These remnants can impose additional constraints on inflationary models and present future observational opportunities in terms of $ \Delta N_{\rm eff} $ \cite{Planck:2018vyg}. Such relics appear in various contexts including supersymmetric models like moduli fields or gravitino \cite{Giudice:1999yt}, and as gravitational waves \cite{Nakayama:2018ptw,Huang:2019lgd,Barman:2023ymn,Chakraborty:2023ocr,Kanemura:2023pnv,Bernal:2023wus,Tokareva:2023mrt,Choi:2024ilx}. Even particles interacting purely gravitationally can be significant \cite{Ema:2016hlw,Kaneta:2022gug,Kolb:2023ydq}.

In this work, we focus on the direct coupling between the inflaton and pseudo-Nambu-Goldstone bosons (pNGBs). The existence of pNGBs is a common prediction in many models beyond the Standard Model (BSM), which often feature spontaneously broken global symmetries at lower energies \cite{Nambu:1984pp,Goldstone:1961eq,Goldstone:1962es}. A prime example is the axion \cite{Weinberg:1977ma,Wilczek:1977pj} which may arise from the spontaneous breaking of the $U(1)_{\rm PQ}$ symmetry \cite{Peccei:1977hh,Peccei:1977ur}. (For recent reviews of axion theories and cosmology, see \cite{Reece:2023czb,OHare:2024nmr}.)

A unique characteristic of pNGBs is their shift symmetry, making them weakly interacting. Although the inflaton field also respects shift symmetry during inflation, this symmetry is broken during reheating. Depending on the charge of the pNGB, we expect couplings at higher dimensions in the following forms:
\begin{align}
\mathcal{L}_{\rm int, 5} = \frac{1}{\Lambda} \phi (\partial \chi)^{2}, && \mathcal{L}_{\rm int, 6} = \frac{1}{\Lambda^{2}} \phi^{2} (\partial \chi)^{2}
\end{align}
where $\phi$ is the inflaton field, $\chi$ is the pNGB, and $\Lambda$ is an unknown cut-off scale.

As a definite application of our general results, we consider a model where the inflaton is the radial mode of the Peccei-Quinn (PQ) scalar field, with a large non-minimal coupling to gravity \cite{Fairbairn:2014zta}. This scenario, which we call PQ inflation, provides a concrete example where the interplay between the inflaton and pNGBs can be studied in detail. In PQ inflation, the PQ scalar field drives inflation and its radial mode acts as the inflaton, which then decays into axions during reheating. This model demonstrates how non-thermal relics can arise from specific inflaton-pNGB interactions and the constraints they impose on the reheating temperature. (See Ref.~\cite{Lee:2023dtw} for a similar consideration.)

This paper is organized as follows. In Section~\ref{section:Inflaton Decay to pNGB during the Reheating}, we consider the inflaton field $\phi$ as a classical background field to account for its coherent nature and derive general formulas for the decay rates to $\chi$ fields from the interactions $\frac{1}{\Lambda} \phi (\partial \chi)^{2}$ and $\frac{1}{\Lambda^{2}} \phi^{2} (\partial \chi)^{2}$ for monomial potentials $V(\phi) \propto \phi^{m}$ with $m = (2, 4, 6)$. In Section~\ref{section:pNGB Abundance}, we apply these results to determine the pNGB abundance by solving the Boltzmann equation for both constant and field-dependent couplings. In Section~\ref{section:Applications}, we examine an inflation model where the PQ radial mode acts as the inflaton field, aided by large gravitational non-minimal coupling, and quantify the non-thermal axion relic produced during reheating, assessing constraints on the model from the $\Delta N_{\rm eff}$ bound. Finally, we conclude in Section~\ref{section:Conclusion}.

\section{Inflaton Decay to pNGB during the Reheating}
\label{section:Inflaton Decay to pNGB during the Reheating}

%
The inflation and reheating stages are characterized by the dynamics of a coherently oscillating inflaton field $\phi$ during the reheating phase following inflation.
Due to its large occupation number, this coherent field configuration can be treated as a classical field \cite{Linde:1990flp}.
Depending on the potential $V(\phi)$, the field evolves according to the Klein-Gordon equation in an expanding universe:
\begin{align}
\ddot{\phi} + 3 H \dot{\phi} + \frac{dV}{d\phi} = 0,
\label{eq:KG}
\end{align}
where $H \equiv \dot{a}/a$ is the Hubble parameter with a dot being derivative with respect to time $ t $.
Broadly speaking, the Hubble friction term causes the amplitude of the field $\phi$ to gradually decrease (slow mode), while the $dV/d\phi$ term induces fast oscillations in the inflaton field (fast mode).
In this study, we assume that the potential minimum during reheating can be approximated as a monomial function $V \propto \phi^{m}$, specifically:
\begin{align}
V(\phi) = \begin{dcases}
\frac{m_{\phi}^{2}}{2} \phi^{2} & (m=2) \\
\frac{\lambda}{4} \phi^{4} & (m=4) \\
\frac{\kappa}{6} \phi^{6} & (m=6)
\end{dcases}
\end{align}
where $m_{\phi}$ is the mass of the field $\phi$, and $\lambda$ and $\kappa$ are self-coupling constants. Note that $\kappa$ has a mass dimension of $[\kappa] = -2$.
For definite examples, the field $\phi$ has approximate solutions of the form:
\begin{align}
\phi(t) \simeq
\begin{dcases}
\phi_{0} \cos (m_{\phi} t) & (m=2) \\
\phi_{0} \, \text{cd} \left( \sqrt{\frac{\lambda \phi_{0}^{2}}{2}} t, -1 \right) & (m=4)
\end{dcases}
\end{align}
where $\phi_{0}$ is the overall amplitude (envelope) of the field $\phi$, and `$\text{cd}$' is one of the Jacobi elliptic functions.
There is no explicit analytic solution for $m=6$.
Neglecting the expansion of the universe (equivalent to setting the second term of Eq.~\eqref{eq:KG} to zero), $\phi_{0}$ remains constant but generally decreases slowly with the universe's expansion, which we refer to as the slow mode.
For later convenience, we introduce the effective mass parameters:
\begin{align}
        (m_{\phi}^{\rm eff})^{2}  \equiv \left. \frac{\partial^{2} V}{\partial \phi^{2}} \right\vert_{\phi=\phi_{0}} = \begin{dcases}
        m_{\phi}^{2}  & (m=2) \\
        3 \lambda \phi_{0}^{2} & (m=4) \\
        5 \kappa \phi_{0}^{4} & (m=6)
    \end{dcases}.
\end{align}
We also decompose the field as:
\begin{align}
    \phi(t) = \sum_{n = - \infty}^{\infty} \phi_{n} e^{- i n \omega t}.
\end{align}
where $ \omega \equiv 2\pi / T $ is the leading, fundamental frequency with the period $T$  \cite{Shtanov:1994ce,Ichikawa:2008ne}.
Explicitly,
\begin{align}
    \omega =  m_{\phi}^{\rm eff}  \times \begin{dcases}
        1 & (m=2) \\
        \frac{1}{2} \sqrt{\frac{\pi}{6}} \frac{\Gamma(3/4)}{\Gamma(5/4)} & (m=4) \\
        \frac{1}{2} \sqrt{\frac{\pi}{15}} \frac{\Gamma(2/3)}{\Gamma(7/6)} & (m=6)
    \end{dcases}.
\end{align}
For instance, in the $m=2$ case, $\phi_{\pm 1} = 1/2$ and zero otherwise. In the $m=4$ case,
\begin{align}
   \phi_{n} =  
   \begin{dcases}
         \frac{\sqrt{\pi} \Gamma(3/4)}{\Gamma(5/4)} 
       \frac{e^{-n \pi/2}}{1+e^{- n \pi}}  & (n~\text{odd}) \\
        0  & (n~\text{even})
   \end{dcases} && (m=4).
\end{align}
In what follows, we will derive the decay rates of $\phi$ to $\chi$, $\Gamma_{\phi \rightarrow \chi}$, from the $\phi (\partial \chi)^{2}$ and $\phi^{2} (\partial \chi)^{2}$ couplings by treating the $\phi$ field as a classical, external field.
%

\subsection{$ \phi (\partial \chi)^{2} $ Coupling}

%
As an explicit starting point, let us consider the following interaction:
\begin{align}
    \mathcal{L}_{\rm int} = - g  \phi (\partial \chi)^{2},
    \label{eq:coupling5}
\end{align}
where the coupling $g$ has mass dimension $[g] = -1$.
Treating $\phi$ as an external current, this can be interpreted as an interaction term in the Hamiltonian:
\begin{align}
    V(t) = g \phi(t) \int d^{3}\vec{x} \,  (\partial \chi)^{2}.
\end{align}
The production rate of the $\chi$ field is then given by:
\begin{align}
    \Gamma = \frac{g^{2}}{16 \pi}   \omega^{4}  \sum_{n=1}^{\infty}  n^{4}\vert \phi_{n} \vert^{2}  = \frac{g^{2}}{32\pi} \langle \ddot{\phi}^{2} \rangle  \label{eq:chi_production_rate}
\end{align}
where $ \langle \, \cdot \, \rangle $ denotes the time average. See Appendix~\ref{appendix:Born Approximation} for details of the calculation.
By comparing the energy loss of the inflaton field $\phi$ and the energy gain of the $\chi$ field, i.e., $\rho_{\phi} \Gamma_{\phi \rightarrow \chi} \Delta t = E_{\chi} \Gamma \Delta t$ for some infinitesimal time interval $\Delta t$ where $E_{\chi}$ is the mean energy of the two-particle state, the decay rate of the inflaton energy density is given by:
\begin{align}
    \Gamma_{\phi \rightarrow \chi} = \Gamma \frac{E_{\chi}}{\rho_{\phi}}
\end{align}
with
\begin{equation}
    \begin{aligned}
    E_{\chi} & \equiv \frac{ \sum_{n} \int d^{3}\vec{p} \, d^{3} \vec{q} \, \delta(\vec{p} + \vec{q} ) E_{f} \delta(E_{f} - n \omega)  \vert \mathcal{M}_{n} \vert^{2} }{  \sum_{n} \int d^{3}\vec{p} \, d^{3} \vec{q} \, \delta(\vec{p} + \vec{q} ) \delta(E_{f} - n \omega)  \vert \mathcal{M}_{n} \vert^{2}  } \\
    & = \frac{\sum_{n=1}^{\infty} (n \omega)^{5} \vert \phi_{n} \vert^{2}}{\sum_{n=1}^{\infty}  (n \omega)^{4} \vert \phi_{n} \vert^{2}} = \omega
    \begin{dcases}
        1 & (m=2) \\
        1.290 & (m=4) \\
        1.700  & (m=6) \label{eq:mean_energy}
    \end{dcases} 
    \end{aligned}    
\end{equation}
where we used the matrix element for mode $n$:
\begin{align}
      \mathcal{M}_{n}
      = - 4 \pi i g  \phi_{n} p 
\end{align}
for the massless pNGB with momentum $p = \vert \vec{p} \vert$. Additionally,
\begin{align}
    \frac{\langle \ddot \phi^{2} \rangle}{\rho_{\phi}} = 
    \begin{dcases}
     m_{\phi}^{2} & (m=2) \\
     0.365 (m_{\phi}^{\rm eff})^{2} & (m=4) \\
    0.230 (m_{\phi}^{\rm eff})^{2} & (m=6)
    \end{dcases}.
\end{align}
Using these results, we obtain the inflaton decay rate as:
\begin{align}
    \Gamma_{\phi \rightarrow \chi} 
    = \frac{ \mathcal{A}_{m} g^{2} (m_{\phi}^{\rm eff})^{3}}{32\pi}, && 
    \mathcal{A}_{m} \equiv \begin{dcases}
        1 & (m=2) \\
        0.231 & (m=4) \\
        0.130 & (m=6)
    \end{dcases} 
    \label{eq:decay_rate}
\end{align}
where we introduce $\mathcal{A}_{m}$ for later convenience.
For $ g = 1 / f_{\chi} $, the decay rate is given by
\begin{align}
    \Gamma_{\phi \rightarrow \chi} = \frac{m_{\phi}^{3}}{32\pi f_{\chi}^{2}}
\end{align}
in the quadratic case, which is what one would obtain using perturbative QFT calculations.
For $m \geq 4$ cases, there is a suppression compared to naive particle-picture results, as indicated by $\mathcal{A}_{m} < 1$.

\subsection{$ \phi^{2} (\partial \chi)^{2} $ Coupling}

Our previous calculation can be straightforwardly generalized to the case of the coupling
\begin{align}
    \mathcal{L}_{\rm int} = - y \phi^{2} (\partial \chi)^{2}, 
    \label{eq:coupling6}
\end{align}
where the coupling $y$ has mass dimension $[y] = -2$.
First, we introduce $ \zeta_{n} $ parameters as
\begin{align}
    \phi^{2}(t) - \langle \phi^{2} \rangle = \sum_{n=-\infty}^{\infty} \zeta_{n} e^{- i n \omega t}.
\end{align}
where we subtract time-independent $\langle \phi^{2} \rangle$ factor. By replacing $\phi_{n}$ with $ \zeta_{n}$ from Eq.~\eqref{eq:chi_production_rate} and Eq.~\eqref{eq:mean_energy}, we have
\begin{align}
    \Gamma = \frac{y^{2}}{16\pi} \omega^{4} \sum_{n=1}^{\infty} n^{4}  \vert  \zeta_{n} \vert^{2}.
\end{align}
The mean energy $E_{\chi}$ is given by:
\begin{align}
    E_{\chi} = \frac{\sum_{n=1}^{\infty} (n \omega)^{5} \vert \zeta_{n} \vert^{2}}{\sum_{n=1}^{\infty} (n \omega)^{4} \vert \zeta_{n} \vert^{2}}  
    = \omega \begin{dcases}
        1 & (m=2) \\
        1.007 & (m=4) \\
        1.019 & (m=6)
    \end{dcases}.
\end{align}
This results in
\begin{align}
    \Gamma_{\phi \rightarrow 
    \chi}
    = \mathcal{B}_{m} \frac{y^{2}}{2\pi}  m_{\phi}^{\rm eff} \rho_{\phi}, && \mathcal{B}_{m} = \begin{dcases}
        1 & (m=2) \\
        1.237 & (m=4) \\
        0.355 & (m=6)
    \end{dcases}. 
\end{align}

\section{PNGB Abundance
}
\label{section:pNGB Abundance}

The energy density of the inflaton field is described by the Boltzmann equation\footnote{Notice that this choice of the definition for $\Gamma_{\phi\to{\rm all}}$ corresponds to introducing the dissipation term in the equation of motion by $\ddot\phi+(3H+\Gamma_{\phi\to{\rm all}}/(1+w_\phi))\dot\phi +dV/d\phi=0$. See Ref.~\cite{Garcia:2020wiy} for comparison. 
We choose this definition to make the computation of the inflaton decay rate simpler.}
\begin{align}
    \dot{\rho}_{\phi}  + 3 H (1 + w_{\phi} ) \rho_{\phi} \simeq - 
    \Gamma_{\phi \rightarrow \text{all}} \rho_{\phi}
\end{align}
where $w_{\phi}$ is the effective equation of state given by
\begin{align}
    w_{\phi} \equiv \frac{ \langle \rho_{\phi} \rangle }{ \langle p_{\phi} \rangle }  =  \frac{m-2}{m+2} = \begin{dcases}
        0 & (m=2) \\
        1/3 & (m=4) \\
        1/2 & (m=6)
    \end{dcases}.
\end{align}
based on the virial theorem, $ \langle \dot \phi^{2} \rangle = m \langle V \rangle $. Here, $ \Gamma_{\phi \rightarrow \text{all}} $ is the total decay rate of the inflaton field.
In this paper, we assume that reheating ends when the Hubble rate at that time becomes comparable to the total decay rate, i.e., $\Gamma_{\phi \rightarrow \text{all}} \simeq H$. 
%

%
During the earlier stages of reheating, with $H \gg \Gamma_{\phi \rightarrow \text{all}}$, we can neglect the decay term, and the energy density follows a simple power law
\begin{align}
    \rho_{\phi} = \rho_{e} \left( \frac{a}{a_{e}} \right)^{-3(1+w_{\phi})}
\end{align}
where $\rho_{e}$ and $a_{e}$ are the energy density of the inflaton field and the scale factor at the end of inflation (i.e., the start of reheating), respectively. This implies $\phi_{0} \propto a^{-6 / (m+2)}$.

Simultaneously, the evolution of the energy density of the relativistic pNGB field $\rho_{\chi}$ is governed by another Boltzmann equation
\begin{align}
    \dot{\rho}_{\chi} + 4 H \rho_{\chi} \simeq 
    \Gamma_{\phi \rightarrow \chi} \rho_{\phi} \label{eq:Boltzmann:pNGB}
\end{align}
where the decay rate of the inflaton to the axion, $\Gamma_{\phi \rightarrow \chi}$, has been obtained in Section~\ref{section:Inflaton Decay to pNGB during the Reheating}.
We neglect the backreaction of the $\chi$ field on the dynamics of the $\phi$ field, which is valid as long as $\rho_{\chi} \ll \rho_{\phi}$.
In this section, we consider the amount of non-thermal, relativistic pNGB remnant for each coupling $\phi (\partial \chi)^{2}$ and $\phi^{2} (\partial \chi)^{2}$.
The results depend significantly on whether we assume constant coupling or field-dependent coupling.
Specifically, we consider cases where $ g \propto \phi_{0}^{-1} $ in Eq.~\eqref{eq:coupling5} and $ y \propto \phi_{0}^{-2} $ in Eq.~\eqref{eq:coupling6}.
Also, it is convenient to change the variable from cosmological time $t$ to the scale factor $a$.
In terms of $a$, the left-hand side of Eq.~\eqref{eq:Boltzmann:pNGB} is rewritten as
\begin{align}
    \dot{\rho}_{\chi} + 4 H \rho_{\chi} = \frac{1}{a^{4}} \frac{d}{dt} (a^{4} \rho_{\chi}) = H_{e}  \frac{1}{a^{3}} \left( \frac{a_{e}}{a} \right)^{3(1+w_{\phi})/2}  \frac{d}{da} (a^{4} \rho_{\chi}) 
\end{align}
where we used $ H = H_{e} \left( a_{e} / a \right)^{3(1+w_{\phi})/2}  $ with $H_{e}$ being the Hubble parameter at the end of the inflation.

\subsection{Constant Coupling}

From the results given in Eq.~\eqref{eq:decay_rate}, Eq.~\eqref{eq:Boltzmann:pNGB} can be easily integrated to provide the following solutions for each potential with $g \phi (\partial \chi )^{2}$ and $y \phi^{2} (\partial \chi )^{2}$ assuming constant coupling coefficients $g$ and $y$:
\begin{itemize}
    \item $ \phi (\partial \chi)^{2} $ Coupling
    
\begin{align}
    \rho_{\chi} =
    \begin{dcases}
        \frac{2}{5} \frac{  g^{2} m_{\phi}^{3}}{32\pi} \frac{\rho_{e}}{H_{e}} \left[ \left( \frac{a_{e}}{a} \right)^{3/2} -   \left( \frac{a_{e}}{a} \right)^{4} \right]  & (m=2) \\
        \frac{ \mathcal{A}_{4} g^{2} (3 \lambda)^{3/2} \phi_{e}^{3}}{32\pi} \frac{\rho_{e}}{H_{e}} \left[ \left( \frac{a_{e}}{a} \right)^{4} - \left( \frac{a_{e}}{a} \right)^{5} \right]  & (m=4) \\
        \frac{  \mathcal{A}_{6} g^{2} (5 \kappa)^{3/2} \phi_{e}^{6}}{88\pi} \frac{\rho_{e}}{H_{e}} \left[ \left( \frac{a_{e}}{a} \right)^{4} - \left( \frac{a_{e}}{a} \right)^{\frac{27}{4}} \right]  & (m=6)
    \end{dcases}
    \label{eq:results_constant_three}
\end{align}

\item $ \phi^{2} (\partial \chi)^{2} $ Coupling

\begin{align}
    \rho_{\chi} = \begin{dcases}
        \frac{  y^{2} m_{\phi}}{\pi} \frac{\rho_{e}^{2}}{H_{e}} \left[ \left( \frac{a_{e}}{a} \right)^{4} -   \left( \frac{a_{e}}{a} \right)^{9/2} \right] & (m=2) \\
        \frac{y^{2}}{2\sqrt{3}\pi} \mathcal{B}_{4} \frac{ \rho_{e}^{2} \phi_{e} \sqrt{\lambda} }{H_{e}} \left[ \left( \frac{a_{e}}{a} \right)^{4} - \left( \frac{a_{e}}{a} \right)^{7} \right] & (m=4) \\
        \frac{2 \sqrt{5} y^{2}}{17\pi} \mathcal{B}_{6} \frac{ \rho_{e}^{2} \phi_{e}^{2} \sqrt{\kappa} }{H_{e}}  \left[ \left( \frac{a_{e}}{a} \right)^{4} - \left( \frac{a_{e}}{a} \right)^{\frac{33}{4}} \right]  & (m=6)
    \end{dcases}
    \label{eq:results_constant_four}
\end{align}

\end{itemize}

Note that only the $\phi (\partial \chi)^{2}$ coupling with $m=2$ decreases slower than pure dilution $\propto a^{-4}$ at large $a$, indicating that the energy density of $\chi$ is dominated by contributions from later times. In this case, the final results are less sensitive to the early dynamics of reheating, such as the non-perturbative preheating stage.

\subsection{Field-dependent Coupling}

As we will see in the application below, some UV models require us to consider the possibility of having field-dependent coupling. Specifically, we consider the couplings given by either $g = \mathcal{C} \phi_{0}^{-1}$ or $y = \mathcal{C} \phi_{0}^{-2}$, with some constant $\mathcal{C}$.
In these cases, the parametric dependence of the results for the two couplings is the same, but the coefficients differ. The results are as follows:
\begin{align}
    \rho_{\chi} 
    =  
    \begin{dcases}
          \frac{2}{11} \frac{  \mathcal{C}^{2} m_{\phi}^{5}}{8\pi} \frac{1}{H_{e}} \left[ \left( \frac{a}{a_{e}} \right)^{3/2} -   \left( \frac{a_{e}}{a} \right)^{4} \right] & (m=2) \\
          \frac{\sqrt{3}\mathcal{C}^{2} \sqrt{\lambda} \rho_{e}^{2} }{2\pi \phi_{e}^{3} H_{e}} \left\{  \frac{3}{4} \mathcal{A}_{4}, \mathcal{B}_{4}  \right\}    \left[ \left( \frac{a_{e}}{a} \right)^{3} -   \left( \frac{a_{e}}{a} \right)^{4} \right]  & (m=4) \\
          \frac{2\sqrt{5}}{5} \frac{\mathcal{C}^{2} \sqrt{\kappa} \rho_{e}^{2} }{\pi \phi_{e}^{2} H_{e}} \left\{  \frac{15}{8} \mathcal{A}_{6}, \mathcal{B}_{6}  \right\}    \left[ \left( \frac{a_{e}}{a} \right)^{4} -   \left( \frac{a_{e}}{a} \right)^{21/4} \right] & (m=6)
    \end{dcases}.
    \label{eq:results_field_dep}
\end{align}
In the above equations, the results for the $\phi (\partial \chi)^{2}$ coupling are given first, followed by those for the $\phi^{2} (\partial \chi)^{2}$ coupling in parentheses, except for the $m=2$ case, where the two cases yield the same answer.

\section{Application: PQ Inflation with Large Non-minimal Coupling}
\label{section:Applications}

As a definite example, let us consider the case where inflaton is assumed to be the radial mode $ \varphi $ of $U(1)_{\rm PQ}$ scalar $ \Phi = \frac{1}{\sqrt{2}} \varphi e^{i \theta} $ with the Lagrangian
\begin{align}
    \frac{ \mathcal{L} }{ \sqrt{-g} } & = (\partial \Phi)^{2} + \xi R \vert \Phi \vert^{2} - \lambda \left( \vert \Phi \vert^{2} - \frac{f_{\chi}^{2}}{2} \right)^{2}
    \\
    & = \frac{1}{2} (\partial \varphi)^{2} + \frac{1}{2} \varphi^{2} (\partial \theta)^{2} + \frac{1}{2} \xi R \varphi^{2} - \frac{\lambda}{4} (\varphi^{2} - f_{\chi}^{2})^{2}
\end{align}
where $ R $ is the Ricci scalar, $\xi $ is the gravitational non-minimal coupling of $\Phi$, $\lambda$ is the quartic coupling of PQ field, and $ f_{\chi} $ is the axion decay constant which sets vev of the radial mode $ \varphi $ \cite{Fairbairn:2014zta}.
Later, we will canonically normalize $ \theta $ field to $ \chi$ field, which corresponds to `axion' field.
In this example, we are interested in the large non-minimal coupling $ \xi \gg 1 $ as one of the most simplest model which fit to the current observations \cite{BICEP:2021xfz,Planck:2018jri,Park:2008hz,Cheong:2021kyc}.%
\footnote{Large non-minimal coupling to the gravity is one of the main feature of the Higgs inflation model \cite{Bezrukov:2007ep,Cheong:2021vdb}.
This model shares similar inflation and reheating dynamics \cite{Bezrukov:2008ut,Garcia-Bellido:2008ycs,Lee:2020yaj}, while allowing large vev $f_{\chi}$ while SM Higgs always have tiny vev compared to inflation/reheating scales.}\footnote{In this work, we are mainly concerning the perturbative regime of the reheating.
However, we note that the earlier preheating stage of the inflation with large non-minimal coupling may be more violent \cite{Ema:2016dny}, which also arouse unitarity issue \cite{Burgess:2009ea,Barbon:2009ya,Burgess:2010zq,Bezrukov:2010jz,Ito:2021ssc}.
In this work, we are focusing on the later stages of the reheating, implicitly assume that preheating stage does not largely modifies the later stages of the inflaton dynamics due to backreaction. This may be accomplished by strong dynamics beyond the unitarity bound or other UV completions of the model.
Many works are done in the Higgs inflation context \cite{Giudice:2010ka,Ema:2017rqn,He:2018mgb,Hamada:2020kuy,Cheong:2019vzl,Lee:2021rzy,Lee:2023wdm}.}
See Appendix~\ref{Appendix:Inflation with Large-Non-minimal Coupling} for the review of the inflation with large gravitational non-minimal coupling.
We will call this inflation model as `PQ inflation' in this paper.
This model also has advantages of suppressing the axion isocurvature \cite{Fairbairn:2014zta} opening new window of high scale inflation consistent to the axion dark matter isocurvature bound \cite{Planck:2018jri}, or in the perspective of reducing axion quality problem \cite{Hamaguchi:2021mmt,Cheong:2022ikv}.
For the rest of the section, we will mainly concern about the amount of the axion relic from direct decay of the inflaton field. This would be left as a relativistic degree of freedom in later time, which is constrained by $\Delta N_{\rm eff}$ measurement. The current constraints $ \Delta N_{\rm eff} \lesssim 0.2$ \cite{Planck:2018vyg}.
In the case of the large non-minimal coupling, it is useful to work in the Einstein frame where the non-minimal coupling is removed by the field redefinitions of the metric $g_{\mu\nu}$ and the field $\varphi$.
In particular, Einstein frame field $ \phi $ is related to $\varphi$ using the following relation \cite{Bezrukov:2008ut}:
\begin{align}
    \frac{d\phi}{d\varphi} = \sqrt{ \frac{\Omega^{2} + 6 \xi^{2} \varphi^{2} / M_{P}^{2}}{ \Omega^{4} } }, &&
     \Omega^{2} \equiv  \frac{M_{P}^{2} + \xi ( \varphi^{2} - f_{\chi}^{2} )}{M_{P}^{2}}.
     \label{eq:canonicalize}
\end{align}
Integrating above relation, the field can be approximated as
\begin{align}
    \phi \simeq 
    \begin{dcases}
        \varphi & \left( \varphi \lesssim \sqrt{\frac{2}{3}} \frac{M_{P}}{\xi}  \right)   \\ 
        \sqrt{ \frac{3}{2} } \frac{\xi \varphi^{2}}{M_{P}}   & \left( \sqrt{\frac{2}{3}} \frac{M_{P}}{\xi} \lesssim \varphi \ll   \frac{M_{P}}{\sqrt{\xi}}
        \right)  
    \end{dcases}
\end{align}
and the potential in the Einstein frame also get corrected from the conformal factor as
\begin{align}
    V_{E}(\phi) = \frac{\lambda}{4 \Omega^{4}} \left[ \varphi(\phi)^{2} - f_{\chi}^{2} \right]^{2}.
\end{align}
For the regime which is relevant for the reheating with $ \phi \ll M_{P} $, it suffices to take $ \Omega \simeq 1 $.
On the other hand, the remaining factors depend on the hierarchy between $ f_{\chi} $ and $ M_{P} / \xi $, while $ f_{\chi} \leq M_{P} / \sqrt{\xi} $ is always assumed to guarantee the positivity of the coefficient of the Ricci scalar.
As reviewed in Appendix~\ref{Appendix:Inflation with Large-Non-minimal Coupling}, Planck measurement of the scalar amplitude $A_{s}$ of the primordial perturbation \cite{Planck:2018jri} dictates us an normalization condition $ \xi^{2} / \lambda \simeq 2.5 \times 10^{9} $ which we also assume to hold.

\subsection{Case I: $  M_{P} / \xi  <  f_{\chi} <  M_{P} / \sqrt{\xi} $}

When $f_{\chi}$ is larger than $ M_{P} / \xi $, the vev of the $\phi$ field is large enough so that we can approximate the potential to be quadratic in terms of the field $ \phi $ as
\begin{align}
   V_{E} (\phi) \simeq 
        \frac{\lambda M_{P}^{2}}{6 \xi^{2}}  \left( \phi - \sqrt{\frac{3}{2}} \frac{\xi f_{\chi}^{2}}{M_{P}} \right)^{2}. 
\end{align}
At first, we can neglect the vacuum expectation value (vev) of the inflaton field $ \phi_{\rm vev} = \sqrt{\frac{3}{2}} \frac{\xi f_{\chi}^{2}}{M_{P}} $ but in the end $ \phi $ stabilizes to $ \phi_{\rm vev} $.
Note that, while the vev of Jordan frame field $ \varphi $ is $f_{\chi}$, going to Einstein field changes the field value of the vev different from $f_{\chi}$.
Therefore, we can approximate the behavior of the inflaton field $ \phi $ as
\begin{align}
    \phi \simeq \begin{dcases}
         \phi_{0} \cos ( m_{\phi} t) &  (\phi_{0} > \phi_{\rm vev}) \\
         \phi_{\rm vev}   & (\phi_{0} < \phi_{\rm vev}) 
    \end{dcases}.
\end{align}
where $ m_{\phi} = \sqrt{ \frac{\lambda M_{P}^{2}}{3 \xi^{2}} } $, and slowly time varying envelop $ \phi_{0} = \phi_{e} \left( a/ a_{e} \right)^{-3/2} $. Also, from the Lagrangian
\begin{align}
     \mathcal{L} \ni \frac{1}{2} \varphi^{2} (\partial \theta)^{2} \simeq  \frac{1}{2} \sqrt{ \frac{3}{2} } \frac{M_{P}}{\xi} \phi  (\partial \theta)^{2},
\end{align}
we normalize $ \theta $ field by defining
\begin{align}
    \chi \equiv 
    \left( \sqrt{ \frac{3}{2} } \frac{M_{P}}{\xi}  \right)^{1/2} \theta \times  \begin{dcases}
         \phi_{0}^{1/2}  & (\phi_{0} \geq \phi_{\rm vev}) \\
         \phi_{\rm vev}^{1/2} & (\phi_{0} < \phi_{\rm vev}). 
    \end{dcases} 
\end{align}
Here, we define $ a_{\rm tr} $ with the condition $ \phi_{0}(a_{\rm tr}) = \phi_{\rm vev} $, hence $  a_{\rm tr} = a_{e} \left( \phi_{e} / \phi_{\rm vev} \right)^{2/3} $. The deformation of the potential and its coupling to the radial mode due to the non-minimal gravitational coupling, even at the vev, is a novel feature that arises when considering a large axion decay constant, $ f_{\chi} > M_{P}/\xi $. In the case of Higgs inflation, the electroweak scale is always much smaller than $ M_{P} / \xi $, so we do not expect significant modifications from the non-minimal coupling near the vacuum.

In this way, we can divide this situation into two stages and identify each stage to one of the cases discussed in previous section as follows:
\begin{itemize}
    \item Stage 1 ($a \leq a_{\rm tr}$): quadratic potential, $ \phi(\partial \chi)^{2} $ with  field dependent coupling $ g = (2 \phi_{0})^{-1} $. [Eq.~\eqref{eq:results_field_dep} with $m=2$]
    \item Stage 2 ($a > a_{\rm tr}$): quadratic potential, $ \phi (\partial \chi)^{2} $ with  constant coupling $ g = (2 \phi_{\rm vev})^{-1} $. [Eq.~\eqref{eq:results_constant_three} with $m=2$]
\end{itemize}
Then, we can derive the energy density of $ \chi $ field as
    \begin{align}
  \rho_{\chi}  (a) \simeq  \begin{dcases}
  \frac{  m_{\phi}^{5}}{176\pi} \frac{1}{H_{e}} \left[ \left( \frac{a}{a_{e}} \right)^{3/2} - \left( \frac{a_{e}}{a} \right)^{4}  \right]&  (a \leq a_{\rm tr}) \\
   \frac{ m_{\phi}^{3}}{320\pi \phi_{\rm vev}^{2}} \frac{ \rho_{\phi,\text{tr}} }{H_{\rm tr}} \left[ \left( \frac{a_{\rm tr}}{a} \right)^{3/2} -   \left( \frac{a_{\rm tr}}{a} \right)^{4} \right] + \rho_{\chi,\text{tr}} \left( \frac{a_{\rm tr}}{a} \right)^{4} & (a>a_{\rm tr})
  \end{dcases}
\end{align}
where we have replaced $ \rho_{e} \rightarrow \rho_{\phi,\text{tr}} \equiv \rho_{\phi}(a_{\rm tr}) $ and $ H_{e} \rightarrow H_{\rm tr} \equiv H(a_{\rm tr}) $ for the $a>a_{\rm tr}$, and the last term with $ \rho_{\chi,\text{tr}} = \rho_{\chi}(a_{\rm tr}) $ is also added to match the solution in the first line at $ a = a_{\rm tr} $.

\begin{figure*}[t]
    \centering
\includegraphics[width=0.45\textwidth]{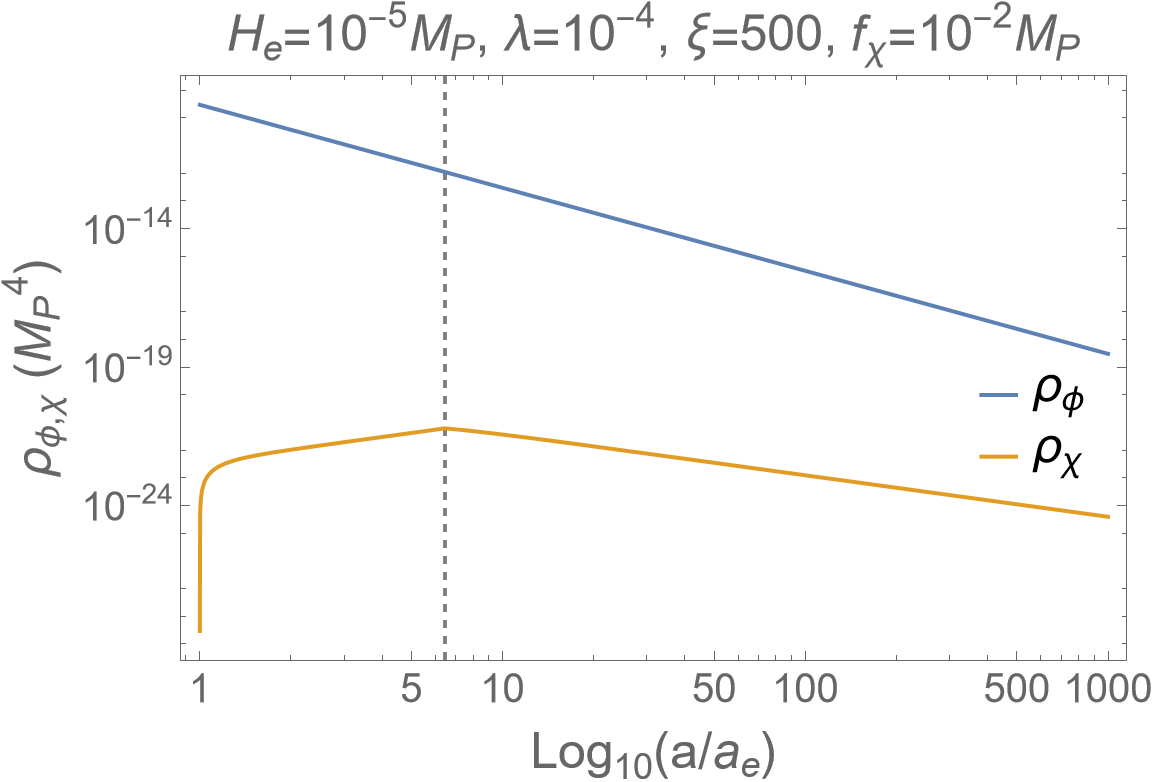}
    \hspace{0.5cm}
\includegraphics[width=0.45\textwidth]{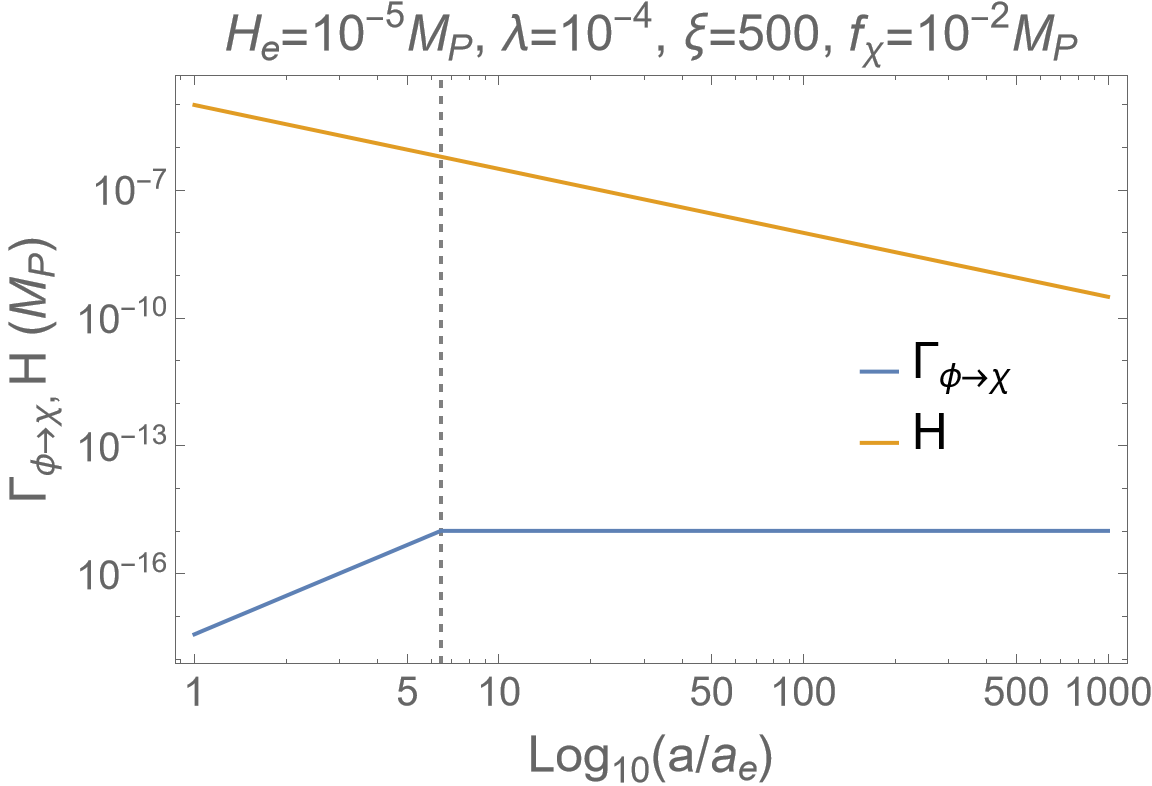}
    \caption{An example of the Case I: $  M_{P} / \xi  <  f_{\chi} <  M_{P} / \sqrt{\xi} $. We took $ H_{e} = 10^{-5} M_{P}$, $ \lambda = 10^{-4}$, $ \xi = 500 $ and $ f_{\chi} = 10^{-2} M_{P} $.
    Here, the vertical dotted line corresponds to $a=a_{\rm tr}$. (Left) Typical evolution of the energy densities of the inflaton ($\rho_{\phi}$, colored in orange) and axion ($\rho_{\chi}$, colored in blue) for the PQ inflation model with large gravitational non-minimal coupling. (Right) Decay rate for each stage (blue) compared to the Hubble rate (orange).}
    \label{fig:case1}
\end{figure*}

Figure~\ref{fig:case1} depicts typical evolution of the energy densities of the inflaton field and the axion field, and the decay rate.
Here we choose $ H_{e} = 10^{-5} M_{P}$, $ \lambda = 10^{-4}$, $ \xi = 500 $ and $ f_{\chi} = 10^{-2} M_{P} $ for an illustration.
While the inflaton energy density decreases like a matter, i.e. $\rho_{\phi} \propto a^{-3}$ through whole history, the energy density of the axion increases at $ a< a_{\rm tr}$ and decreases at $ a > a_{\rm tr}$ but slower than the inflaton energy density.
%

\subsection{Case II: $ 0 <  f_{\chi} < M_{P} / \xi $ }

%
On the other hand, when $f_{\chi}$ is smaller than $M_{P} / \xi$, we have the potential in the Einstein frame in the form of
\begin{align}
    V_{E}(\phi) \simeq \begin{dcases}
       \frac{\lambda M_{P}^{2}}{6 \xi^{2}}  \phi^{2}   & \left( \sqrt{ \frac{2}{3} } \frac{M_{P}}{\xi} < \phi \ll  \sqrt{\frac{3}{2}} M_{P}    \right) \\
      \frac{\lambda}{4} (\phi^{2} - f_{\chi}^{2})^{2} & \left( \phi < \sqrt{\frac{2}{3}} \frac{M_{P}}{\xi}  \right)   
    \end{dcases}
\end{align}
and $\phi$ stabilizes to $ \varphi_{\rm vev} = f_{\chi}  $ in the end.
The first transition from the quadratic potential to quartic one happens at
\begin{align}
    \sqrt{\frac{2}{3}}  \frac{M_{P}}{\xi} \equiv  \phi_{\rm tr1} = \phi_{e} \left( \frac{a_{\rm tr1}}{a_{e}} \right)^{-3/2} 
\end{align}
and the second one from the quartic to quadratic (near the vev) happens at
\begin{align}
   f_{\chi} \equiv \phi_{\rm tr2} = \phi_{\rm tr 1} \left( \frac{a_{\rm tr2}}{a_{\rm tr1}} \right)^{-1}.
\end{align}
Then, in a similar fashion to the previous case, we can divide into three stages as
\begin{itemize}
    \item Stage 1 ($a \leq a_{\rm tr1}$): quadratic potential, $ \phi(\partial \chi)^{2} $ with  field dependent coupling $ g = (2 \phi_{0})^{-1} $. [Eq.~\eqref{eq:results_field_dep} with $m=2$]
    \item Stage 2 ($a_{\rm tr1} < a \leq a_{\rm tr2}$): quartic potential, $ \phi^{2} (\partial \chi)^{2} $ with  field dependent coupling $ y = 1 / (2 \phi_{0}^{2}) $. [Eq.~\eqref{eq:results_field_dep} with $m=4$]
    \item Stage 3 ($ a > a_{\rm tr2} $): quadratic potential, $ \phi^{2} (\partial \chi)^{2} $ with  constant coupling $ y = 1 / (2 f_{\chi}^{2}) $. [Eq.~\eqref{eq:results_constant_four} with $m=2$]
\end{itemize}
Also, the energy density of the daughter particles are given as
    \begin{align}
    \rho_{\chi}^{\rm} (a) \simeq
    \begin{dcases}
        \frac{ m_{\phi}^{5}}{176\pi} \frac{1}{H_{e}} \left[ \left( \frac{a}{a_{e}} \right)^{3/2} - \left( \frac{a_{e}}{a} \right)^{4}  \right] & (a \leq a_{\rm tr1})   \\
        \frac{ \sqrt{3 \lambda} \rho_{\phi,\text{tr1}}^{2} }{ 8 \pi \phi_{\rm tr 1}^{3} H_{\rm tr 1}}  \mathcal{B}_{4} 
    \left[ \left( \frac{a_{\rm tr1}}{a} \right)^{3} - \left( \frac{a_{\rm tr1}}{a} \right)^{4} \right] + \rho_{\chi,\text{tr1}} \left( \frac{ a_{\rm tr1} }{ a } \right)^{4} & (a_{\rm tr1}  < a \leq a_{\rm tr2}) \\
    \frac{ \sqrt{3 \lambda} \rho_{\phi,\text{tr2}}^{2} \phi_{\rm tr2} }{18\pi f_{\chi}^{4} H_{\rm tr2}} \mathcal{B}_{4} \left[ \left( \frac{a_{e}}{a} \right)^{4} - \left( \frac{a_{e}}{a} \right)^{7} \right] +
    \rho_{\chi,\text{tr2}} \left( \frac{ a_{\rm tr2} }{ a } \right)^{4} & (a > a_{\rm tr2})
    \end{dcases} 
\end{align} 
where we introduced $ \rho_{\phi,\text{tr1(2)}} \equiv \rho_{\phi} (a_{\rm tr1(2)}) $, $ \rho_{\chi,\text{tr1(2)}} \equiv \rho_{\chi} (a_{\rm tr1(2)}) $, $ H_{\rm tr1(2)} \equiv H(a_{\rm tr1(2)})$ and added last terms for last two cases for the proper matching.

\begin{figure*}[t]
    \centering
\includegraphics[width=0.45\textwidth]{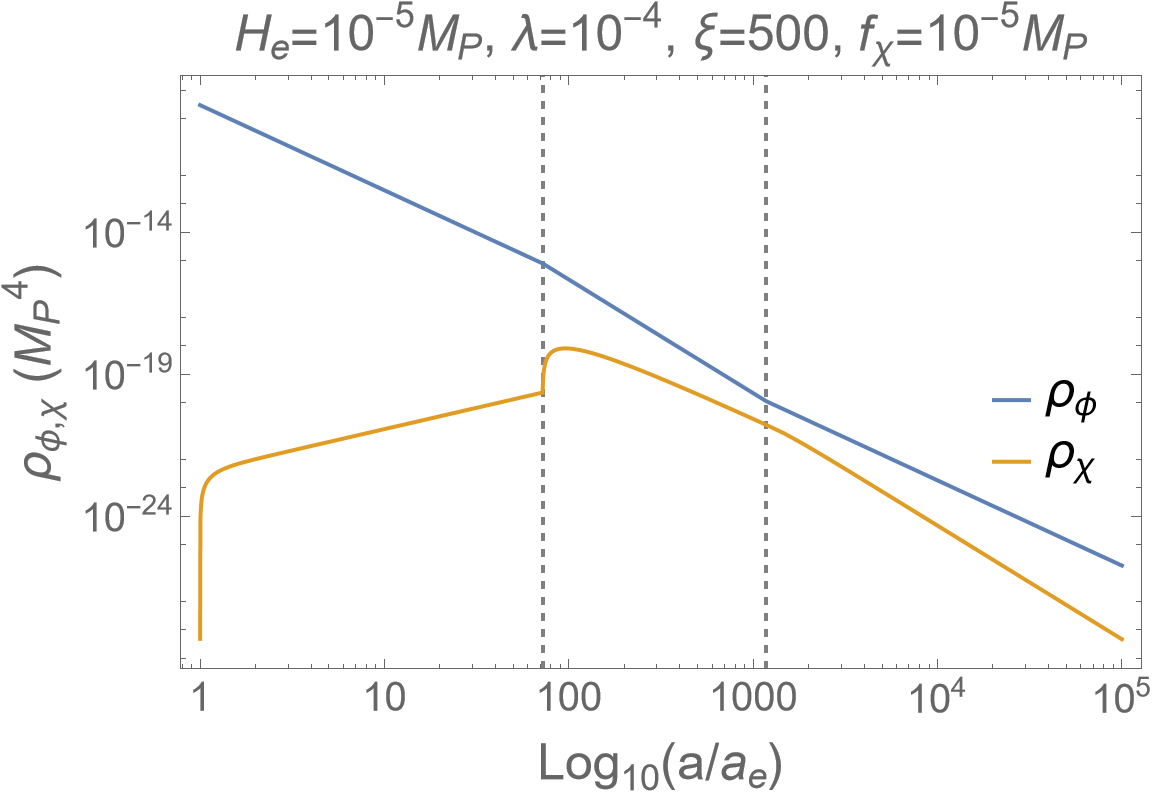}
    \hspace{0.5cm}
\includegraphics[width=0.45\textwidth]{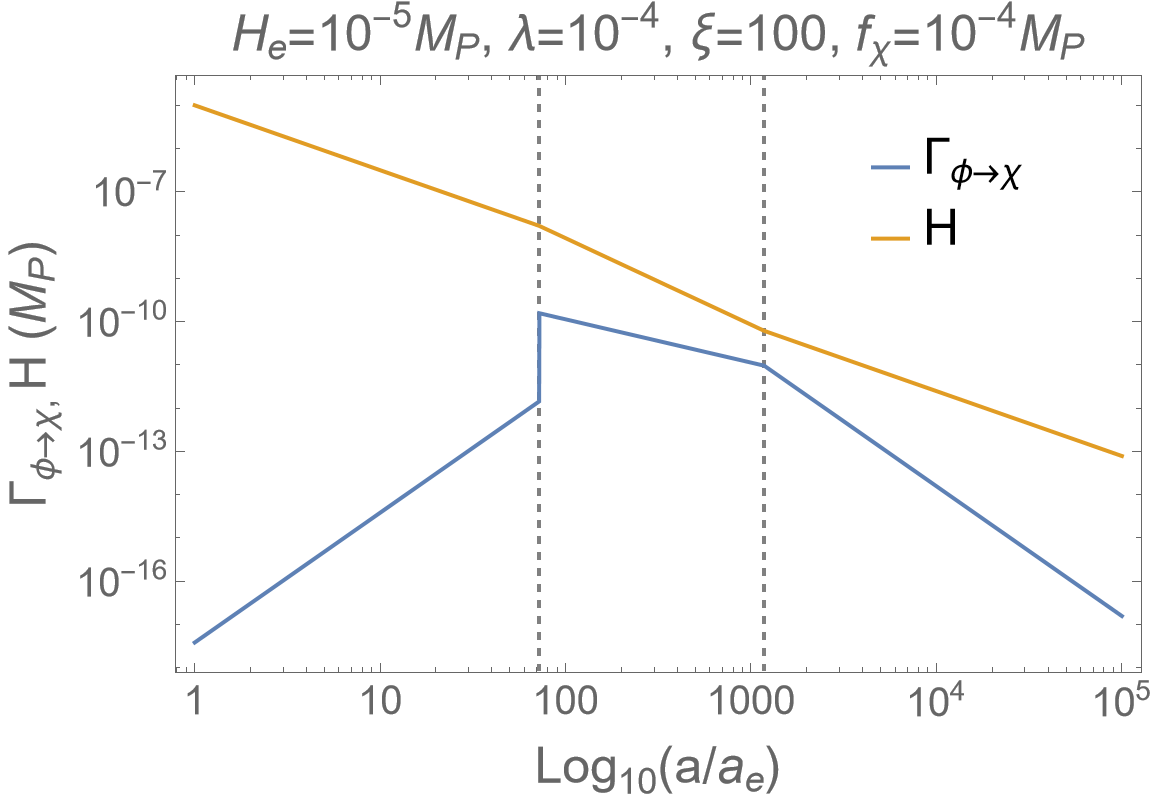}
    \caption{An example of the Case II: $  f_{\chi} <   M_{P} / \xi   $. We took $ H_{e} = 10^{-5} M_{P}$, $ \lambda = 10^{-4}$, $ \xi = 500 $ and $ f_{\chi} = 10^{-4} M_{P} $.
    Here, the vertical dotted lines correspond to $a=a_{\rm tr1 }$ and $a=a_{\rm tr2}$ from the left to right. (Left) Typical evolution of the energy densities of the inflaton ($\rho_{\phi}$, colored in orange) and axion ($\rho_{\chi}$, colored in blue) for the PQ inflation model with large gravitational non-minimal coupling.  (Right) Decay rate for each stage (blue) compared to the Hubble rate (orange). Sudden increase of the decay rate in the second stage after $ a = a_{\rm tr1}$ explains enhancement of the energy density which is observed in the left panel.}
    \label{fig:case2}
\end{figure*}

The Figure~\ref{fig:case2} exemplifies the evolution of the energy densities of the inflaton field and the axion field with $ H_{e} = 10^{-5} M_{P}$, $ \lambda = 10^{-4}$, $ \xi = 500 $ and $ f_{\chi} = 10^{-5} M_{P} $.
In this case the inflaton energy density decreases like a matter ($\rho_{\phi} \propto a^{-3}$) first at $ a<a_{\rm tr1}$ and behaves like a radiation ($\rho_{\phi} \propto a^{-4}$) for $a_{\rm tr1}
< a < a_{\rm tr2}$ and then become matter like again.
Axion energy density increases at $ a< a_{\rm tr1}$ and decreases at $ a > a_{\rm tr1}$.

The reheating should end earlier than the axion energy density dominates over the inflaton energy density: $ \rho_{\chi} < \rho_{\phi} $ always. 
If $\rho_\chi\simeq \rho_\phi$ before the end of reheating, we cannot neglect the backreaction of the axion product to the dynamics of the inflaton field. Moreover, a large amount of the relativistic axion field at the time of the end of the reheating would be ruled out from $ \Delta N_{\rm eff} $ bound \cite{Planck:2018vyg}. More explicitly, we will impose the condition
\begin{align}
    \frac{\rho_{\chi}(a_{\rm reh})}{\rho_{r}(a_{\rm reh})} < 0.10 \left( 
 \frac{\Delta N_{\rm eff}}{0.3}\right) \left(\frac{g_*(T_{\rm reh})}{106.75}\right)^{1/3}.
\end{align}
where $ \rho_{r}$ is the energy density of the total radiation and $a_{\rm reh}$ is determined implicitly by the other decay channel of the inflaton field, which is assumed to dominate. See Appendix~\ref{Appendix:DeltaNeff} for the derivation of this bound. Also, we assume the the energy density of the inflaton corresponds to the energy density of the background radiation other than the relativistic axion with the reheating temperature $T_{\rm reh}$, i.e.
\begin{align}
    \rho_{\phi}(a_{\rm reh}) = \rho_{r}(a_{\rm reh}) =\frac{\pi^{2}}{30} g_{*}(T_{\rm reh}) T_{\rm reh}^{4}
\end{align}
where we take $ g_{*} (T_{\rm reh}) = 106.25 $ in this work and assume that the total energy density of the inflaton background switch to the radiation at the end of the reheating in the first equality.

\begin{figure}[t]
    \centering
\includegraphics[width=0.6\textwidth]{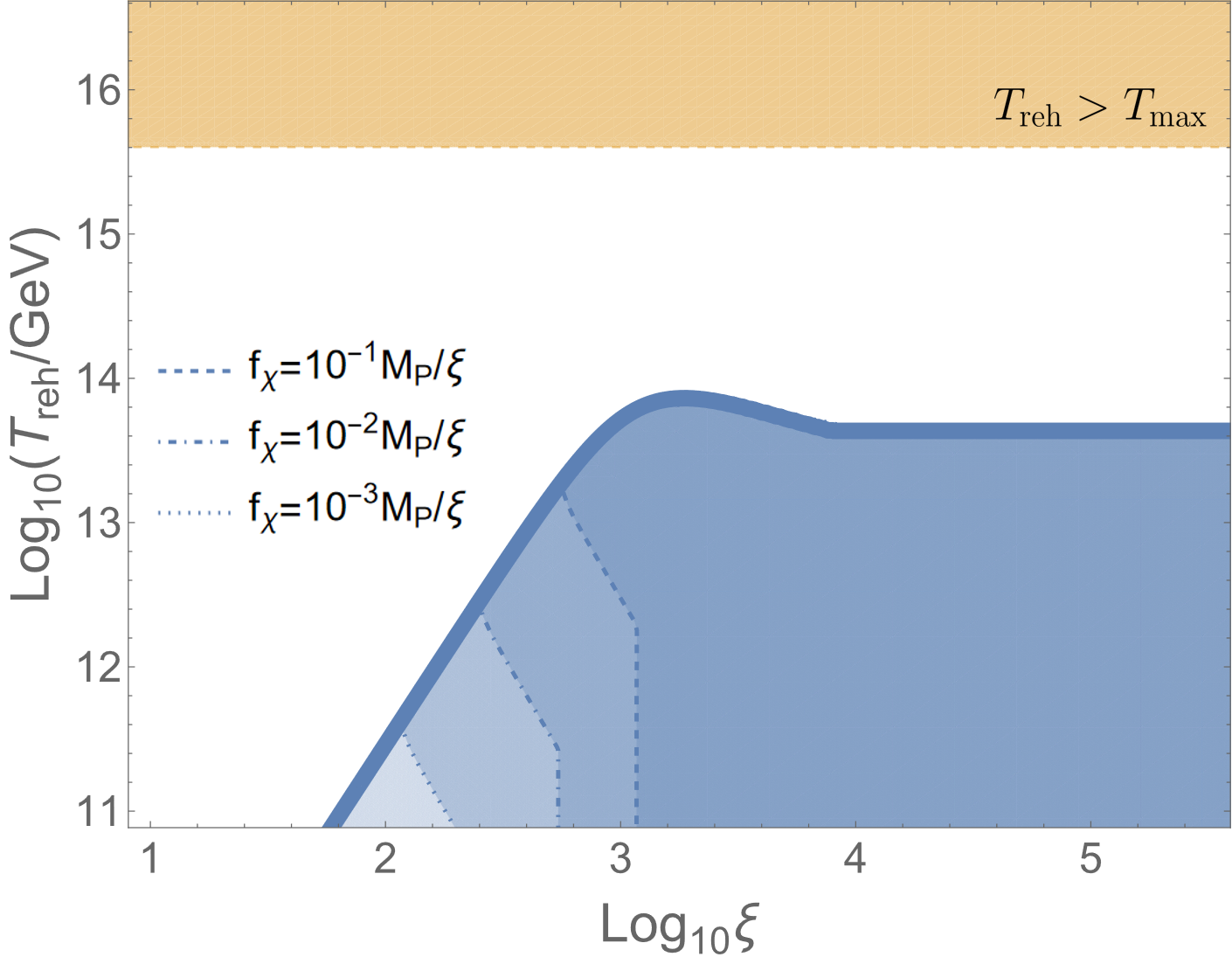}
    \caption{Constraint on the reheating temperature for PQ inflation. Blue regions are excluded from the $ \Delta N_{\rm eff}$ bound coming from the relativistic axion generated during the reheating depending on the ratio between $f_{\chi}$ and $ M_{P} / \xi$. Dashed, dot-dashed, dotted lines represent $ f_{\chi} = (10^{-1}, \, 10^{-2}, \, 10^{-3}) M_{P} / \xi $ respectively and thick solid line corresponds to $ f_{\chi} \rightarrow 0$ limit. Orange regions are also not viable because the reheating temperature exceed the maximal value of instantaneous reheating.}
\label{fig:reheating}
\end{figure}

In Figure~\ref{fig:reheating}, we plot the constraint on the reheating temperature $T_{\rm reh}$ for PQ inflation coming from the above argument. For a given $ \xi $, we choose $ \lambda $ to satisfy the CMB normalization condition, $\xi^{2} / \lambda \simeq 2.5 \times 10^{9}$. In the plot, blue colored regions are excluded from the $ \Delta N_{\rm eff}$ bound. Dashed, dot-dashed, dotted lines represent $ f_{\chi} = (10^{-1}, \, 10^{-2}, \, 10^{-3}) M_{P} / \xi $ respectively and thick solid line corresponds to $ f_{\chi} \rightarrow 0$ limit. For instance, in the case of $ \xi \gtrsim 10^{4} $, there is lower bound of the reheating temperature so that $ T_{\rm reh} \gtrsim 4 \times 10^{13} \, \text{GeV} $, while for $ f_{\chi} \rightarrow 0 $ limit, smaller $ \xi \lesssim 10^{3}$ case has a weaker lower bound so that $ T_{\rm reh}  \gtrsim 3 \times 10^{6} \cdot \xi^{5/2} \, \text{GeV} $. Orange regions are also not viable because the reheating temperature exceed the maximal value of instantaneous reheating $ T_{\rm max} \simeq 4 \times 10^{15} \, \text{GeV}$.

\section{Conclusion}
\label{section:Conclusion}

In this paper, we have investigated the potential implications of the reheating phase following inflation, focusing specifically on the decay of the inflaton field into pNGBs, including axions. We derived general formulas for the decay rates of inflaton to pNGBs under both constant and field-dependent coupling scenarios, assuming a monomial potential $ V(\phi) \propto \phi^{m} $ with $ m = (2,4,6)$.

Our analysis revealed that the presence of pNGBs during reheating, which appear universally in many BSM models, can lead to the production of non-thermal relics. These relics may manifest as dark radiation and influence the effective number of neutrino species $ \Delta N_{\rm eff} $, providing potential observational signatures. This connection offers opportunities to detect BSM remnants from the reheating phase in the early universe.

We applied our general results to a specific model where the inflaton is identified with the radial mode of the PQ scalar field, which possesses a large non-minimal coupling to gravity $ \xi$. Our findings indicate that the bounds on $ \Delta N_{\rm eff} $ impose significant constraints on the reheating temperature, particularly when the axion decay constant $ f_{\chi}$ is smaller than $ M_{P} / \xi $.

Our work highlights that the reheating stage, often considered a black box, may have rich phenomenological consequences associated with the coupling between inflaton and pNGBs.

\acknowledgments

SML thanks to Dhong Yeon Cheong, Koichi Hamaguchi, Yoshiki Kanazawa, Natsumi Nagata and Seong Chan Park for many insights and discussions from the related works. The work of SML is supported by Samsung Science Technology Foundation under Project Number SSTF-BA2302-05 and
the National Research Foundation of Korea (NRF) Grant RS-2023-00211732. This work was supported by JSPS KAKENHI Grant Number 23K17691~(TT) and MEXT KAKENHI 23H04515~(TT).

\appendix

\section{Born Approximation}
\label{appendix:Born Approximation}

In this section, we present the details of the calculation of the decay rate using the Born approximation, which we omitted in the main text.

As a definite case, let us consider the following interaction first:
\begin{align}
    \mathcal{L}_{\rm int} = - g  \phi (\partial \chi)^{2}.
\end{align}
Here, we treat $ \phi (t)$ as a classical field, and only promote $ \chi $ field as a quantum field $ \hat{\chi} $:
\begin{align}
   \hat{\chi} = \int \frac{ d^{3} \vec{p} }{(2\pi)^{3/2} \sqrt{2E_{p}} } \left( 
 e^{ip x} a_{\vec{p}} + e^{-ip x} a_{\vec{p}}^{\dagger} \right).
\end{align}
For the brevity, we will omit hat hereafter. In the Hamiltonian, this gives the interaction
\begin{align}
    V(t) = g \phi(t) \int d^{3}\vec{x} \,  (\partial \chi)^{2}.
\end{align}
From this interaction, what we are interested in is the transition amplitude from the vacuum $ \vert 0 \rangle $ to the final state with two $ \chi$s with momentum $ \vec{k}_{1}$ and $ \vec{k}_{2}$, i.e. $\vert \vec{k}_{1}, \vec{k}_{2} \rangle$:
\begin{align}
    & \langle \vec{k}_{1}, \vec{k}_{2} \vert g \phi(t) (\partial_{\mu} \chi) (\partial^{\mu} \chi) \vert 0 \rangle \\
& = - g  \int d^{4}x \,  \frac{1}{ (2\pi)^{3} \sqrt{E_{k_{1}} E_{k_{2}} } } \sum_{n = -\infty}^{\infty} \phi_{n} \left( E_{k_{1}} E_{k_{2}} - \vec{k}_{1} \cdot \vec{k}_{2}  \right) e^{i (E_{k_{1}} + E_{k_{2}} - n \omega )t} e^{ i (\vec{k}_{1} + \vec{k}_{2} ) \cdot \vec{x}}   \\
& = - 2 \pi g  \delta(\vec{k}_{1} + \vec{k}_{2})  \sum_{n = -\infty}^{\infty}  \phi_{n} \frac{  E_{k_{1}}^{2} + \vert \vec{k}_{1} \vert^{2}  }{E_{k_{1}}} \delta (2 E_{k_{1}} - n \omega )
\end{align}

Then the production rate of $ \chi $ field becomes
\begin{equation}
    \begin{aligned}
    \Gamma & = \frac{1}{2}  \frac{1}{2\pi} (2 \pi g)^{2}  \sum_{n=1}^{\infty} \vert \phi_{n} \vert^{2} \int \frac{d^{3}k}{(2\pi)^{3}}  \delta ( 2E_{k} - \omega ) \frac{ 
 (E_{k}^{2} + \vert \vec{k} \vert^{2} )^{2}  }{E_{k}^{2}} \\
    & =  4 \pi g^{2} \sum_{n=1}^{\infty} \vert \phi_{n} \vert^{2} \int \frac{ 4\pi k^{2} \, dk}{(2\pi)^{3}} \delta (2k-n \omega) \cdot k^{2}  \\
    & = \frac{g^{2}}{16 \pi}   \omega^{4}  \sum_{n=1}^{\infty}  n^{4}\vert \phi_{n} \vert^{2}  = \frac{g^{2}}{32\pi} \langle \ddot{\phi}^{2} \rangle 
\end{aligned}
\end{equation}
This corresponds to Eq.~\eqref{eq:chi_production_rate} in the main text. The results for the $ \mathcal{L}_{\rm int} = - y  \phi^{2} (\partial \chi)^{2} $ case can be obtained using the same steps.

\section{Inflation with Large-Non-minimal Coupling}

\label{Appendix:Inflation with Large-Non-minimal Coupling}

In this section, we summarize the essential results of inflation with large gravitational non-minimal coupling \cite{Bezrukov:2007ep,Rubio:2018ogq,Cheong:2021kyc}, which we use in Section~\ref{section:Applications}. The main purpose is to clarify many details used in the inflation model.

Let us first consider the Lagrangian in Jordan frame with field $ \varphi$
\begin{align}
   S = \int d^{4} \sqrt{-g_{J}} \left[ -\frac{M^{2} + \xi \varphi^{2}}{2} R_{J} + \frac{1}{2} (\partial \varphi)^{2} - V(\varphi) \right]
\end{align}
with $ M^{2} \equiv M_{P}^{2} - \xi f_{\chi}^{2}$ and $ g_{J} \equiv \det g_{J \mu\nu} $.

We redefine the metric
\begin{align}
    g_{\mu\nu} \equiv \Omega^{2} g_{J\mu\nu}, && \Omega^{2} \equiv  \frac{M_{P}^{2} + \xi ( \varphi^{2} - f_{\chi}^{2} )}{M_{P}^{2}}. 
\end{align}
Then, the transformed action becomes
\begin{align}
    S = \int d^{4} \sqrt{-g} \left[ -\frac{M_{P}}{2} R + \frac{1}{2} \Pi(\varphi) (\partial \varphi)^{2} - \frac{V(\varphi)}{\Omega^{4}} \right]
\end{align}
where
\begin{align}
    \Pi (\varphi) \equiv \frac{\Omega^{2} + 6 \xi^{2} \varphi^{2} / M_{P}^{2}}{ \Omega^{4} }
\end{align}
motivating us to introduce the canonicalized field $ d \phi / d \varphi = \sqrt{\Pi(\varphi)}$. This corresponds to Eq.~\eqref{eq:canonicalize}. This expression can be integrated analytically \cite{Garcia-Bellido:2008ycs,Rubio:2018ogq}. Here, instead of presenting full expressions, we provide approximated ones with comparison to the exact one:
\begin{align}
    \phi \simeq 
    \begin{dcases}
        \varphi & \left( \varphi \lesssim \sqrt{\frac{2}{3}} \frac{M_{P}}{\xi}  \right)   \\ 
        \sqrt{ \frac{3}{2} } \frac{\xi \varphi^{2}}{M_{P}}   & \left( \sqrt{\frac{2}{3}} \frac{M_{P}}{\xi} \lesssim \varphi \ll   \frac{M_{P}}{\sqrt{\xi}} 
        \right)     \\ 
       6 M_{P} \log \left( \frac{\sqrt{\xi} \varphi}{M_{P}} \right) & \left( \varphi \gg \frac{M_{P}}{\sqrt{\xi}} \right)
    \end{dcases}
\end{align}

During the inflationary regime with large field value, this implies that
\begin{align}
    V_{E} ( \phi ) \simeq \frac{\lambda M_{P}^{4}}{4 \xi^{2}} \left[ 1 - \exp \left( - \sqrt{ \frac{2}{3} } \frac{\phi}{M_{P}} \right) \right]^{2}.
\end{align}
The end of inflation is defined when one of the slow-roll parameters, $ \epsilon_{V} \simeq \frac{M_{P}^{2}}{2} \left( \frac{V_{E}^{\prime}}{V_{E}} \right)^{2}, $ becomes unity, providing the field value at the end of inflation, $ \phi_{e} \simeq 0.94 M_{P} $. This is also regarded as the initial field value at the beginning of the reheating stage.

CMB pivot scale corresponds to where the expansion happens about $ \log{(a_{e} / a_{*})} \equiv N_{e} \simeq 50-60 $ where $a_{i}$ is the scale factor of the universe when the modes corresponding to CMB observations leave the horizon. In terms of the potential, it is approximated as
\begin{align}
 N_{e} \simeq \int_{\phi_{*}}^{\phi_{e}} \frac{1}{\sqrt{2\epsilon_{V}}} \frac{d\phi}{M_{P}}
\end{align}
implying $ \phi_{*} \simeq 5M_{P} $.

The observational result on the amplitude of the scalar perturbations $ A_{s} \simeq 2.1 \times 10^{-9} $ \cite{Planck:2018jri} can be interpreted as a bound on the scale of inflation
\begin{align}
    A_{s} \simeq \frac{H_{\rm inf}}{8\pi^{2} \epsilon_{V*} M_{P}^{2}} 
\end{align}
where $ \epsilon_{V*} \equiv \epsilon_{V}(\phi_{*}) \simeq 3 / (4N_{e}^{2}) $. This dictates the normalization $ \xi^{2} / \lambda \simeq 2.5 \times 10^{9} $ \cite{Bezrukov:2007ep}. This condition is used throughout the main text.

\section{$ \Delta N_{\rm eff} $ Bound}

\label{Appendix:DeltaNeff}

When there exists an extra relativistic degree of freedom $X$, this would also add extra energy density $ \rho_{X} $. This amount is usually parameterized by $ \Delta N_{\rm eff} $, as the ratio with respect to the energy density of the neutrinos,
\begin{align}
    \Delta N_{\rm eff} \equiv \frac{\rho_{X}}{\rho_{\nu}}.
\end{align}
In terms of the energy density of the photon, $ \rho_{X} $ can be rewritten as
\begin{align}
\rho_{X}(a_{0}) = \Delta N_{\text{eff}} \cdot \frac{7}{8}\left(\frac{4}{11}\right)^{4/3}   \rho_{\gamma}(a_{0})
\end{align}
with $a_{0}$ being scale factor of today. The Planck constraints on $ \Delta N_{\rm eff}$ is $\Delta N_{\rm eff} \lesssim 0.3$ \cite{Planck:2018vyg}.

The goal of this appendix is to derive the bound on the energy density at the time of the reheating from $ \Delta N_{\rm eff}$ bound. In the main text, we specify $ X $ as $ \chi $.
 
\begin{align}
\frac{\rho_{X}(a_{0})}{\rho_{\gamma}(a_{0})} = 
\frac{\rho_{X}(a_{\rm reh})}{\rho_{r}(a_{\rm reh})}
\cdot
\frac{\rho_{X}(a_{0})/\rho_{X}(a_{\rm reh})}{\rho_{r}(a_{0})/\rho_{r}(a_{\rm reh})}
\cdot
\frac{\rho_{r}(a_{0})}{\rho_{\gamma}(a_{0})} 
\end{align}
where $ \rho_{r} $ is the energy density of the total radiation, i.e. $ \rho_{r} = \rho_{\gamma} + \rho_{\nu} + \rho_{X}$ although the contribution from $ \rho_{X}$ is negligible. Here, assuming that $ X $ decouples all other degrees of freedom, $ \rho_{X} \propto a^{-4} $, while the total radiation receive corrections by having different effective number of relativistic degrees of freedom $g_{*}(T)$ as the temperature decreases:
\begin{align}
\rho_{r} = \frac{\pi^{2}}{30} g_{*}(T) T^{4}.
\end{align}
On the other hand, from the entropy conservation, we have $ g_{*s}(T)T^{3}a^{3}=\text{const}$. Then, we have 
\begin{align}
\frac{\rho_{X}(a_{0})/\rho_{X}(a_{\rm reh})}{\rho_{r}(a_{0})/\rho_{r}(a_{\rm reh})}
=
\frac{g_{*}(T_{\text{reh}})^{-1/3}}{g_*(T_0) g_{*s}(T_0)^{-4/3}}.
\end{align}
where we assumed $g_*(T_{\rm reh}) = g_{*s}(T_{\rm reh})$ at high temperature.

Finally, this implies that
\begin{align}
\frac{\rho_{X}(a_{\rm reh})}{\rho_{r}(a_{\rm reh})} & = \frac{\rho_{X}(a_{0})}{\rho_{\gamma}(a_{0})} \left( \frac{\rho_{r}(a_{0})}{\rho_{\gamma}(a_{0})} 
 \right)^{-1} \left( \frac{g_{*}(T_{\text{reh}})^{-1/3}}{g_*(T_0) g_{*s}(T_0)^{-4/3}} \right)^{-1} \\
 & = \Delta N_{\text{eff}} \cdot \frac{7}{8}\left(\frac{4}{11}\right)^{4/3} \left( \frac{\rho_{\gamma}(a_{0})}{\rho_{r}(a_{0})} 
 \right) \left( \frac{g_*(T_0) }{g_{*s}(T_0)^{4/3}}  \right) g_{*}(T_{\text{reh}})^{1/3} \\
 & \simeq 0.10  \left( \frac{\Delta N_{\rm eff}}{0.3} \right) \left(\frac{g_*(T_{\rm reh})}{106.75}\right)^{1/3}
\end{align}
where we used $ g_{*}(T_{0}) = 3.38 $, $ g_{*s}(T_{0}) = 3.94 $ in the last line.

\bibliographystyle{JHEP}
\bibliography{pNGB}

\providecommand{\href}[2]{#2}\begingroup\raggedright\begin{thebibliography}{10}

\bibitem{Linde:1981mu}
A.D.~Linde, \emph{{A New Inflationary Universe Scenario: A Possible Solution of the Horizon, Flatness, Homogeneity, Isotropy and Primordial Monopole Problems}}, \href{https://doi.org/10.1016/0370-2693(82)91219-9}{\emph{Phys. Lett. B} {\bfseries 108} (1982) 389}.

\bibitem{Albrecht:1982mp}
A.~Albrecht, P.J.~Steinhardt, M.S.~Turner and F.~Wilczek, \emph{{Reheating an Inflationary Universe}}, \href{https://doi.org/10.1103/PhysRevLett.48.1437}{\emph{Phys. Rev. Lett.} {\bfseries 48} (1982) 1437}.

\bibitem{Kofman:1994rk}
L.~Kofman, A.D.~Linde and A.A.~Starobinsky, \emph{{Reheating after inflation}}, \href{https://doi.org/10.1103/PhysRevLett.73.3195}{\emph{Phys. Rev. Lett.} {\bfseries 73} (1994) 3195} [\href{https://arxiv.org/abs/hep-th/9405187}{{\ttfamily hep-th/9405187}}].

\bibitem{Kofman:1997yn}
L.~Kofman, A.D.~Linde and A.A.~Starobinsky, \emph{{Towards the theory of reheating after inflation}}, \href{https://doi.org/10.1103/PhysRevD.56.3258}{\emph{Phys. Rev. D} {\bfseries 56} (1997) 3258} [\href{https://arxiv.org/abs/hep-ph/9704452}{{\ttfamily hep-ph/9704452}}].

\bibitem{Lozanov:2019jxc}
K.D.~Lozanov, \emph{{Lectures on Reheating after Inflation}},  \href{https://arxiv.org/abs/1907.04402}{{\ttfamily 1907.04402}}.

\bibitem{Weinberg:2003sw}
S.~Weinberg, \emph{{Adiabatic modes in cosmology}}, \href{https://doi.org/10.1103/PhysRevD.67.123504}{\emph{Phys. Rev. D} {\bfseries 67} (2003) 123504} [\href{https://arxiv.org/abs/astro-ph/0302326}{{\ttfamily astro-ph/0302326}}].

\bibitem{Cook:2015vqa}
J.L.~Cook, E.~Dimastrogiovanni, D.A.~Easson and L.M.~Krauss, \emph{{Reheating predictions in single field inflation}}, \href{https://doi.org/10.1088/1475-7516/2015/04/047}{\emph{JCAP} {\bfseries 04} (2015) 047} [\href{https://arxiv.org/abs/1502.04673}{{\ttfamily 1502.04673}}].

\bibitem{Cheong:2021kyc}
D.Y.~Cheong, S.M.~Lee and S.C.~Park, \emph{{Reheating in models with non-minimal coupling in metric and~Palatini formalisms}}, \href{https://doi.org/10.1088/1475-7516/2022/02/029}{\emph{JCAP} {\bfseries 02} (2022) 029} [\href{https://arxiv.org/abs/2111.00825}{{\ttfamily 2111.00825}}].

\bibitem{Planck:2018vyg}
{\scshape Planck} collaboration, \emph{{Planck 2018 results. VI. Cosmological parameters}}, \href{https://doi.org/10.1051/0004-6361/201833910}{\emph{Astron. Astrophys.} {\bfseries 641} (2020) A6} [\href{https://arxiv.org/abs/1807.06209}{{\ttfamily 1807.06209}}].

\bibitem{Giudice:1999yt}
G.F.~Giudice, I.~Tkachev and A.~Riotto, \emph{{Nonthermal production of dangerous relics in the early universe}}, \href{https://doi.org/10.1088/1126-6708/1999/08/009}{\emph{JHEP} {\bfseries 08} (1999) 009} [\href{https://arxiv.org/abs/hep-ph/9907510}{{\ttfamily hep-ph/9907510}}].

\bibitem{Nakayama:2018ptw}
K.~Nakayama and Y.~Tang, \emph{{Stochastic Gravitational Waves from Particle Origin}}, \href{https://doi.org/10.1016/j.physletb.2018.11.023}{\emph{Phys. Lett. B} {\bfseries 788} (2019) 341} [\href{https://arxiv.org/abs/1810.04975}{{\ttfamily 1810.04975}}].

\bibitem{Huang:2019lgd}
D.~Huang and L.~Yin, \emph{{Stochastic Gravitational Waves from Inflaton Decays}}, \href{https://doi.org/10.1103/PhysRevD.100.043538}{\emph{Phys. Rev. D} {\bfseries 100} (2019) 043538} [\href{https://arxiv.org/abs/1905.08510}{{\ttfamily 1905.08510}}].

\bibitem{Barman:2023ymn}
B.~Barman, N.~Bernal, Y.~Xu and O.~Zapata, \emph{{Gravitational wave from graviton Bremsstrahlung during reheating}}, \href{https://doi.org/10.1088/1475-7516/2023/05/019}{\emph{JCAP} {\bfseries 05} (2023) 019} [\href{https://arxiv.org/abs/2301.11345}{{\ttfamily 2301.11345}}].

\bibitem{Chakraborty:2023ocr}
A.~Chakraborty, M.R.~Haque, D.~Maity and R.~Mondal, \emph{{Inflaton phenomenology via reheating in light of primordial gravitational waves and the latest BICEP/Keck data}}, \href{https://doi.org/10.1103/PhysRevD.108.023515}{\emph{Phys. Rev. D} {\bfseries 108} (2023) 023515} [\href{https://arxiv.org/abs/2304.13637}{{\ttfamily 2304.13637}}].

\bibitem{Kanemura:2023pnv}
S.~Kanemura and K.~Kaneta, \emph{{Gravitational Waves from Particle Decays during Reheating}},  \href{https://arxiv.org/abs/2310.12023}{{\ttfamily 2310.12023}}.

\bibitem{Bernal:2023wus}
N.~Bernal, S.~Cl\'ery, Y.~Mambrini and Y.~Xu, \emph{{Probing reheating with graviton bremsstrahlung}}, \href{https://doi.org/10.1088/1475-7516/2024/01/065}{\emph{JCAP} {\bfseries 01} (2024) 065} [\href{https://arxiv.org/abs/2311.12694}{{\ttfamily 2311.12694}}].

\bibitem{Tokareva:2023mrt}
A.~Tokareva, \emph{{Gravitational waves from inflaton decay and bremsstrahlung}}, \href{https://doi.org/10.1016/j.physletb.2024.138695}{\emph{Phys. Lett. B} {\bfseries 853} (2024) 138695} [\href{https://arxiv.org/abs/2312.16691}{{\ttfamily 2312.16691}}].

\bibitem{Choi:2024ilx}
G.~Choi, W.~Ke and K.A.~Olive, \emph{{Minimal production of prompt gravitational waves during reheating}}, \href{https://doi.org/10.1103/PhysRevD.109.083516}{\emph{Phys. Rev. D} {\bfseries 109} (2024) 083516} [\href{https://arxiv.org/abs/2402.04310}{{\ttfamily 2402.04310}}].

\bibitem{Ema:2016hlw}
Y.~Ema, R.~Jinno, K.~Mukaida and K.~Nakayama, \emph{{Gravitational particle production in oscillating backgrounds and its cosmological implications}}, \href{https://doi.org/10.1103/PhysRevD.94.063517}{\emph{Phys. Rev. D} {\bfseries 94} (2016) 063517} [\href{https://arxiv.org/abs/1604.08898}{{\ttfamily 1604.08898}}].

\bibitem{Kaneta:2022gug}
K.~Kaneta, S.M.~Lee and K.-y.~Oda, \emph{{Boltzmann or Bogoliubov? Approaches compared in gravitational particle production}}, \href{https://doi.org/10.1088/1475-7516/2022/09/018}{\emph{JCAP} {\bfseries 09} (2022) 018} [\href{https://arxiv.org/abs/2206.10929}{{\ttfamily 2206.10929}}].

\bibitem{Kolb:2023ydq}
E.W.~Kolb and A.J.~Long, \emph{{Cosmological gravitational particle production and its implications for cosmological relics}},  \href{https://arxiv.org/abs/2312.09042}{{\ttfamily 2312.09042}}.

\bibitem{Nambu:1984pp}
Y.~Nambu, \emph{{SUPERCONDUCTIVITY AND PARTICLE PHYSICS}}, {\emph{Physica B} {\bfseries 126} (1984) 328}.

\bibitem{Goldstone:1961eq}
J.~Goldstone, \emph{{Field Theories with Superconductor Solutions}}, \href{https://doi.org/10.1007/BF02812722}{\emph{Nuovo Cim.} {\bfseries 19} (1961) 154}.

\bibitem{Goldstone:1962es}
J.~Goldstone, A.~Salam and S.~Weinberg, \emph{{Broken Symmetries}}, \href{https://doi.org/10.1103/PhysRev.127.965}{\emph{Phys. Rev.} {\bfseries 127} (1962) 965}.

\bibitem{Weinberg:1977ma}
S.~Weinberg, \emph{{A New Light Boson?}}, \href{https://doi.org/10.1103/PhysRevLett.40.223}{\emph{Phys. Rev. Lett.} {\bfseries 40} (1978) 223}.

\bibitem{Wilczek:1977pj}
F.~Wilczek, \emph{{Problem of Strong $P$ and $T$ Invariance in the Presence of Instantons}}, \href{https://doi.org/10.1103/PhysRevLett.40.279}{\emph{Phys. Rev. Lett.} {\bfseries 40} (1978) 279}.

\bibitem{Peccei:1977hh}
R.D.~Peccei and H.R.~Quinn, \emph{{CP Conservation in the Presence of Instantons}}, \href{https://doi.org/10.1103/PhysRevLett.38.1440}{\emph{Phys. Rev. Lett.} {\bfseries 38} (1977) 1440}.

\bibitem{Peccei:1977ur}
R.D.~Peccei and H.R.~Quinn, \emph{{Constraints Imposed by CP Conservation in the Presence of Instantons}}, \href{https://doi.org/10.1103/PhysRevD.16.1791}{\emph{Phys. Rev. D} {\bfseries 16} (1977) 1791}.

\bibitem{Reece:2023czb}
M.~Reece, \emph{{TASI Lectures: (No) Global Symmetries to Axion Physics}}, \href{https://doi.org/10.22323/1.439.0008}{\emph{PoS} {\bfseries TASI2022} (2024) 008} [\href{https://arxiv.org/abs/2304.08512}{{\ttfamily 2304.08512}}].

\bibitem{OHare:2024nmr}
C.A.J.~O'Hare, \emph{{Cosmology of axion dark matter}}, \href{https://doi.org/10.22323/1.454.0040}{\emph{PoS} {\bfseries COSMICWISPers} (2024) 040} [\href{https://arxiv.org/abs/2403.17697}{{\ttfamily 2403.17697}}].

\bibitem{Fairbairn:2014zta}
M.~Fairbairn, R.~Hogan and D.J.E.~Marsh, \emph{{Unifying inflation and dark matter with the Peccei-Quinn field: observable axions and observable tensors}}, \href{https://doi.org/10.1103/PhysRevD.91.023509}{\emph{Phys. Rev. D} {\bfseries 91} (2015) 023509} [\href{https://arxiv.org/abs/1410.1752}{{\ttfamily 1410.1752}}].

\bibitem{Lee:2023dtw}
H.M.~Lee, A.G.~Menkara, M.-J.~Seong and J.-H.~Song, \emph{{Peccei-Quinn Inflation at the Pole and Axion Kinetic Misalignment}},  \href{https://arxiv.org/abs/2310.17710}{{\ttfamily 2310.17710}}.

\bibitem{Linde:1990flp}
A.D.~Linde, \emph{{Particle physics and inflationary cosmology}}, vol.~5 (1990), [\href{https://arxiv.org/abs/hep-th/0503203}{{\ttfamily hep-th/0503203}}].

\bibitem{Shtanov:1994ce}
Y.~Shtanov, J.H.~Traschen and R.H.~Brandenberger, \emph{{Universe reheating after inflation}}, \href{https://doi.org/10.1103/PhysRevD.51.5438}{\emph{Phys. Rev. D} {\bfseries 51} (1995) 5438} [\href{https://arxiv.org/abs/hep-ph/9407247}{{\ttfamily hep-ph/9407247}}].

\bibitem{Ichikawa:2008ne}
K.~Ichikawa, T.~Suyama, T.~Takahashi and M.~Yamaguchi, \emph{{Primordial Curvature Fluctuation and Its Non-Gaussianity in Models with Modulated Reheating}}, \href{https://doi.org/10.1103/PhysRevD.78.063545}{\emph{Phys. Rev. D} {\bfseries 78} (2008) 063545} [\href{https://arxiv.org/abs/0807.3988}{{\ttfamily 0807.3988}}].

\bibitem{Garcia:2020wiy}
M.A.G.~Garcia, K.~Kaneta, Y.~Mambrini and K.A.~Olive, \emph{{Inflaton Oscillations and Post-Inflationary Reheating}}, \href{https://doi.org/10.1088/1475-7516/2021/04/012}{\emph{JCAP} {\bfseries 04} (2021) 012} [\href{https://arxiv.org/abs/2012.10756}{{\ttfamily 2012.10756}}].

\bibitem{BICEP:2021xfz}
{\scshape BICEP, Keck} collaboration, \emph{{Improved Constraints on Primordial Gravitational Waves using Planck, WMAP, and BICEP/Keck Observations through the 2018 Observing Season}}, \href{https://doi.org/10.1103/PhysRevLett.127.151301}{\emph{Phys. Rev. Lett.} {\bfseries 127} (2021) 151301} [\href{https://arxiv.org/abs/2110.00483}{{\ttfamily 2110.00483}}].

\bibitem{Planck:2018jri}
{\scshape Planck} collaboration, \emph{{Planck 2018 results. X. Constraints on inflation}}, \href{https://doi.org/10.1051/0004-6361/201833887}{\emph{Astron. Astrophys.} {\bfseries 641} (2020) A10} [\href{https://arxiv.org/abs/1807.06211}{{\ttfamily 1807.06211}}].

\bibitem{Park:2008hz}
S.C.~Park and S.~Yamaguchi, \emph{{Inflation by non-minimal coupling}}, \href{https://doi.org/10.1088/1475-7516/2008/08/009}{\emph{JCAP} {\bfseries 08} (2008) 009} [\href{https://arxiv.org/abs/0801.1722}{{\ttfamily 0801.1722}}].

\bibitem{Bezrukov:2007ep}
F.L.~Bezrukov and M.~Shaposhnikov, \emph{{The Standard Model Higgs boson as the inflaton}}, \href{https://doi.org/10.1016/j.physletb.2007.11.072}{\emph{Phys. Lett. B} {\bfseries 659} (2008) 703} [\href{https://arxiv.org/abs/0710.3755}{{\ttfamily 0710.3755}}].

\bibitem{Cheong:2021vdb}
D.Y.~Cheong, S.M.~Lee and S.C.~Park, \emph{{Progress in Higgs inflation}}, \href{https://doi.org/10.1007/s40042-021-00086-2}{\emph{J. Korean Phys. Soc.} {\bfseries 78} (2021) 897} [\href{https://arxiv.org/abs/2103.00177}{{\ttfamily 2103.00177}}].

\bibitem{Bezrukov:2008ut}
F.~Bezrukov, D.~Gorbunov and M.~Shaposhnikov, \emph{{On initial conditions for the Hot Big Bang}}, \href{https://doi.org/10.1088/1475-7516/2009/06/029}{\emph{JCAP} {\bfseries 06} (2009) 029} [\href{https://arxiv.org/abs/0812.3622}{{\ttfamily 0812.3622}}].

\bibitem{Garcia-Bellido:2008ycs}
J.~Garcia-Bellido, D.G.~Figueroa and J.~Rubio, \emph{{Preheating in the Standard Model with the Higgs-Inflaton coupled to gravity}}, \href{https://doi.org/10.1103/PhysRevD.79.063531}{\emph{Phys. Rev. D} {\bfseries 79} (2009) 063531} [\href{https://arxiv.org/abs/0812.4624}{{\ttfamily 0812.4624}}].

\bibitem{Lee:2020yaj}
S.M.~Lee, K.-y.~Oda and S.C.~Park, \emph{{Spontaneous Leptogenesis in Higgs Inflation}}, \href{https://doi.org/10.1007/JHEP03(2021)083}{\emph{JHEP} {\bfseries 03} (2021) 083} [\href{https://arxiv.org/abs/2010.07563}{{\ttfamily 2010.07563}}].

\bibitem{Ema:2016dny}
Y.~Ema, R.~Jinno, K.~Mukaida and K.~Nakayama, \emph{{Violent Preheating in Inflation with Nonminimal Coupling}}, \href{https://doi.org/10.1088/1475-7516/2017/02/045}{\emph{JCAP} {\bfseries 02} (2017) 045} [\href{https://arxiv.org/abs/1609.05209}{{\ttfamily 1609.05209}}].

\bibitem{Burgess:2009ea}
C.P.~Burgess, H.M.~Lee and M.~Trott, \emph{{Power-counting and the Validity of the Classical Approximation During Inflation}}, \href{https://doi.org/10.1088/1126-6708/2009/09/103}{\emph{JHEP} {\bfseries 09} (2009) 103} [\href{https://arxiv.org/abs/0902.4465}{{\ttfamily 0902.4465}}].

\bibitem{Barbon:2009ya}
J.L.F.~Barbon and J.R.~Espinosa, \emph{{On the Naturalness of Higgs Inflation}}, \href{https://doi.org/10.1103/PhysRevD.79.081302}{\emph{Phys. Rev. D} {\bfseries 79} (2009) 081302} [\href{https://arxiv.org/abs/0903.0355}{{\ttfamily 0903.0355}}].

\bibitem{Burgess:2010zq}
C.P.~Burgess, H.M.~Lee and M.~Trott, \emph{{Comment on Higgs Inflation and Naturalness}}, \href{https://doi.org/10.1007/JHEP07(2010)007}{\emph{JHEP} {\bfseries 07} (2010) 007} [\href{https://arxiv.org/abs/1002.2730}{{\ttfamily 1002.2730}}].

\bibitem{Bezrukov:2010jz}
F.~Bezrukov, A.~Magnin, M.~Shaposhnikov and S.~Sibiryakov, \emph{{Higgs inflation: consistency and generalisations}}, \href{https://doi.org/10.1007/JHEP01(2011)016}{\emph{JHEP} {\bfseries 01} (2011) 016} [\href{https://arxiv.org/abs/1008.5157}{{\ttfamily 1008.5157}}].

\bibitem{Ito:2021ssc}
A.~Ito, W.~Khater and S.~Rasanen, \emph{{Tree-level unitarity in Higgs inflation in the metric and the Palatini formulation}}, \href{https://doi.org/10.1007/JHEP06(2022)164}{\emph{JHEP} {\bfseries 06} (2022) 164} [\href{https://arxiv.org/abs/2111.05621}{{\ttfamily 2111.05621}}].

\bibitem{Giudice:2010ka}
G.F.~Giudice and H.M.~Lee, \emph{{Unitarizing Higgs Inflation}}, \href{https://doi.org/10.1016/j.physletb.2010.10.035}{\emph{Phys. Lett. B} {\bfseries 694} (2011) 294} [\href{https://arxiv.org/abs/1010.1417}{{\ttfamily 1010.1417}}].

\bibitem{Ema:2017rqn}
Y.~Ema, \emph{{Higgs Scalaron Mixed Inflation}}, \href{https://doi.org/10.1016/j.physletb.2017.04.060}{\emph{Phys. Lett. B} {\bfseries 770} (2017) 403} [\href{https://arxiv.org/abs/1701.07665}{{\ttfamily 1701.07665}}].

\bibitem{He:2018mgb}
M.~He, R.~Jinno, K.~Kamada, S.C.~Park, A.A.~Starobinsky and J.~Yokoyama, \emph{{On the violent preheating in the mixed Higgs-$R^2$ inflationary model}}, \href{https://doi.org/10.1016/j.physletb.2019.02.008}{\emph{Phys. Lett. B} {\bfseries 791} (2019) 36} [\href{https://arxiv.org/abs/1812.10099}{{\ttfamily 1812.10099}}].

\bibitem{Hamada:2020kuy}
Y.~Hamada, K.~Kawana and A.~Scherlis, \emph{{On Preheating in Higgs Inflation}}, \href{https://doi.org/10.1088/1475-7516/2021/03/062}{\emph{JCAP} {\bfseries 03} (2021) 062} [\href{https://arxiv.org/abs/2007.04701}{{\ttfamily 2007.04701}}].

\bibitem{Cheong:2019vzl}
D.Y.~Cheong, S.M.~Lee and S.C.~Park, \emph{{Primordial black holes in Higgs-$R^2$ inflation as the whole of dark matter}}, \href{https://doi.org/10.1088/1475-7516/2021/01/032}{\emph{JCAP} {\bfseries 01} (2021) 032} [\href{https://arxiv.org/abs/1912.12032}{{\ttfamily 1912.12032}}].

\bibitem{Lee:2021rzy}
S.M.~Lee, T.~Modak, K.-y.~Oda and T.~Takahashi, \emph{{The $R^2$-Higgs inflation with two Higgs doublets}}, \href{https://doi.org/10.1140/epjc/s10052-021-09978-w}{\emph{Eur. Phys. J. C} {\bfseries 82} (2022) 18} [\href{https://arxiv.org/abs/2108.02383}{{\ttfamily 2108.02383}}].

\bibitem{Lee:2023wdm}
S.M.~Lee, T.~Modak, K.-y.~Oda and T.~Takahashi, \emph{{Ultraviolet sensitivity in Higgs-Starobinsky inflation}}, \href{https://doi.org/10.1088/1475-7516/2023/08/045}{\emph{JCAP} {\bfseries 08} (2023) 045} [\href{https://arxiv.org/abs/2303.09866}{{\ttfamily 2303.09866}}].

\bibitem{Hamaguchi:2021mmt}
K.~Hamaguchi, Y.~Kanazawa and N.~Nagata, \emph{{Axion quality problem alleviated by nonminimal coupling to gravity}}, \href{https://doi.org/10.1103/PhysRevD.105.076008}{\emph{Phys. Rev. D} {\bfseries 105} (2022) 076008} [\href{https://arxiv.org/abs/2108.13245}{{\ttfamily 2108.13245}}].

\bibitem{Cheong:2022ikv}
D.Y.~Cheong, K.~Hamaguchi, Y.~Kanazawa, S.M.~Lee, N.~Nagata and S.C.~Park, \emph{{Axion quality problem and nonminimal gravitational coupling in the Palatini formulation}}, \href{https://doi.org/10.1103/PhysRevD.108.015007}{\emph{Phys. Rev. D} {\bfseries 108} (2023) 015007} [\href{https://arxiv.org/abs/2210.11330}{{\ttfamily 2210.11330}}].

\bibitem{Rubio:2018ogq}
J.~Rubio, \emph{{Higgs inflation}}, \href{https://doi.org/10.3389/fspas.2018.00050}{\emph{Front. Astron. Space Sci.} {\bfseries 5} (2019) 50} [\href{https://arxiv.org/abs/1807.02376}{{\ttfamily 1807.02376}}].

\end{thebibliography}\endgroup

\end{document}